\documentclass[double]{article}
\usepackage{times}
\usepackage{algorithm}
\usepackage{algorithmic}
\usepackage{wrapfig}
\usepackage{verbatim}
\usepackage{amsmath}
\usepackage{graphicx}
\usepackage{subcaption}
\usepackage[active]{srcltx}
\usepackage{color}
\usepackage{lineno}
\usepackage{dirtytalk}
\usepackage{amsthm}
\usepackage{authblk}
\usepackage{listings}
\usepackage{tikz}
\usetikzlibrary{shapes.geometric, arrows}

\usepackage{epsfig,graphicx,amsfonts}
\usepackage{amsmath,amsthm,amssymb}
\usepackage{xcolor}

\usepackage{adjustbox}
\usepackage{multirow}
\usepackage{setspace}
\usepackage{hyperref}
\usepackage{comment}

\long\def\comment #1\commentend{}

\begin{document}
%\documentstyle[amsfonts]{article}

% Following creates the title page
\title{\Large Mathematical Model of Dating Apps' Influence on Sexually Transmitted Diseases Spread}

% \author{Anonlymous for review}
 \author{Teddy Lazebnik$^{1,2,*}$ \\ \(^1\) Department of Mathematics, Ariel University, Ariel, Israel\\ \(^2\) Department of Cancer Biology, Cancer Institute, University College London, London, UK\\ \(^*\) Corresponding author: lazebnik.teddy@gmail.com }

\date{ }

\maketitle 

\begin{abstract}
Sexually transmitted diseases (STDs) are a group of pathogens infecting new hosts through sexual interactions. Due to its social and economic burden, multiple models have been proposed to study the spreading of pathogens. In parallel, in the ever-evolving landscape of digital social interactions, the pervasive utilization of dating apps has become a prominent facet of modern society. Despite the surge in popularity and the profound impact on relationship formation, a crucial gap in the literature persists regarding the potential ramifications of dating apps usage on the dynamics of STDs. In this paper, we address this gap by presenting a novel mathematical framework — an extended Susceptible-Infected-Susceptible (SIS) epidemlazebnik.teddy@gmail.com iological model to elucidate the intricate interplay between dating apps engagement and the propagation of STDs. Namely, as dating apps are designed to make users revisit them and have mainly casual sexual interactions with other users, they increase the number of causal partners, which increases the overall spread of STDS. Using extensive simulation, based on real-world data, explore the effect of dating apps adoption and control on the STD spread. We show that an increased adoption of dating apps can result in an STD outbreak if not handled appropriately. \\ \\
\noindent
\textbf{Keywords:} Sexual behavior dynamics; extended SIS model; multi-pathogen epidemic; digital intervention policy; public health.
\end{abstract}

\maketitle \thispagestyle{empty}

% Begin using page numbers and a header
\pagestyle{myheadings} \markboth{Draft:  \today}{Draft:  \today}
\setcounter{page}{1}% reset page number to 1

\section{Introduction}
\label{sec:introduction}
Sexually transmitted diseases (STDs) are a significant public health challenge, exerting a substantial social and economic burden globally \cite{stds_1,stds_2,stds_3,stds_4}. With an estimated 376 million new infections reported annually, the widespread prevalence of STDs necessitates comprehensive investigations into their transmission dynamics and the factors that contribute to their propagation \cite{who,infectious_diseases}. In Particular, data from the Centers for Disease Control and Prevention (CDC) in the U.S. illustrates a notable upsurge in newly reported cases of chlamydia, gonorrhea, and syphilis since 2013 \cite{intro_1,stds_increase_1,stds_increase_2}.

As part of a larger trend of social interactions moving into the digital world \cite{moving_to_digital_1,moving_to_digital_2}, the rise of online dating platforms has introduced increased complexity and versatility into the way individuals find life and sexual partners \cite{dating_apps_1,digital_complexity}. For instance, recent research has established a correlation between the use of online dating applications and a history of five or more previous sexual partners among young adults \cite{intro_2,dating_apps_increase_partners}. To effectively capture the interplay between sexual network structures, partner formation, and STD transmission, researchers have developed diverse mathematical frameworks \cite{intro_chat_1,intro_chat_2,intro_chat_3}. However, existing models often overlook the inherent heterogeneity in individual-level link formation, as they rely on mean-field approximations at the pair level or statistical characterizations of sexual networks \cite{intro_3,intro_4,intro_5}. 

These efforts, however, have predominantly centered on traditional modes of social interaction, overlooking the transformative impact of digital platforms in reshaping interpersonal connections. In the contemporary landscape, dating apps have emerged as a pervasive and influential feature of modern society, revolutionizing the way individuals initiate and cultivate relationships \cite{dating_apps_2,morrissey2020emerging}. The meteoric rise in dating app adoption and usage underscores the need to reevaluate existing disease transmission models. In this work, we introduce a novel mathematical framework based on an extended Susceptible-Infectious-Susceptible (SIS) epidemiological model to investigate the intricate interplay between dating app usage and STD transmission spread dynamics.

The rest of the paper is organized as follows. Section \ref{sec:related_work} presents an overview of the dating app design, objective, and social impact as well as STD spread dynamics models. In Section \ref{sec:model}, we outline the proposed mathematical model constructed from a graph-based spatial model, the influence of dating apps, an extended multi-pathogen SIS model, and an agent-based simulation implementation to allow heterogeneous population dynamics. Next, in Section \ref{sec:experiment}, we describe the experiment design of the proposed model with a realistic configuration followed by the obtained results from the experiments. Finally, Section \ref{sec:discussion} provides an analysis of the results as well as the strengths and limitations of the proposed model with possible future work. 

\section{Related Work}
\label{sec:related_work}
In order to understand the STD spread dynamics and the role of dating apps in these settings, we overview the current design, objective, and social influence of dating apps on the population followed by a disruption of previous epidemiological models in general and for STDs, in particular.  

\subsection{Dating apps}
\label{sec:dating_apps}
As technology evolved, a greater number of dating apps were created to help individuals find their partner, whether it may be sexual or romantic \cite{dating_apps_origin}. The proliferation of dating apps has ushered in a new era of interpersonal connectivity, revolutionizing the way individuals form relationships and engage in romantic interactions \cite{dating_intro_1,dating_intro_2}. Dating apps have witnessed exceptional growth in recent years, with an increasing number of users engaging in diverse forms of interaction facilitated by these platforms \cite{datting_usage}. Interestingly, the business objective of these apps is usually counter to their declared marketing for users since apps are financially gaining from users using the app as much as possible while promising to help find someone that would make the users leave the app \cite{dating_intro_3,dating_intro_4}.

While studies about the nature of users' objectives in such dating apps are spread across the "hookup" and meaningful relationship line, all agree about the fact these mobile applications increase the amount of romantic and sexual interactions overall \cite{dating_apps_not_hookups,dating_apps_cause_hookups_1,dating_apps_cause_hookups_2}. This fast-paced scenario can fuel an STD spread since the more sexual partners a person has, the higher the likelihood of coming into contact with an infected individual as each new partner represents a potential source of infection, especially if they have multiple partners themselves \cite{more_sex_more_infection}. Hence, dating apps have garnered interest within the realm of public health research \cite{dating_health_problem_1,dating_health_problem_2}. Notably, the potential links between dating app usage and increased sexual risk behavior have raised concerns regarding STDs transmission dynamics \cite{dating_problem}. For example, Miller (2020) \cite{dating_problem_example} surveyed almost a thousand university students who used dating apps in the previous year versus students who did not. The author found that students who used dating apps were statistically more likely to have a greater number of sexual partners during this time but the author was not able to find a statistical increase in STD infection. 

Overall, dating apps operate as a tool for an individual in the population to increase their network of possible casual and long-term sexual relationships. This increase can be integrated into current STD spread models to understand the possible role dating apps play in STD spread dynamics. 

\subsection{STD spread modeling}
\label{sec:std_models}

Mathematical and computational models are key tools for understanding pandemic spread and designing intervention policies that help control a pandemic's spread \cite{sir_review,sir_review2}. In particular, coupled ordinary and partial differential equations, as well as simpler growth-curve equations, are previously used to capture pandemic spread in general \cite{intro_3_cost_plants,intro_related_1_cost_plants,intro_related_2_cost_plants,teddy_review,intro_related_4_cost_plants,journal_1,journal_2} and STD diseases spread, in particular \cite{std_2_2_intro_1,std_2_2_intro_2,std_2_2_intro_3}.

More often than not, the models describing the spread of STDs extend the Susceptible-Infectious-Recovered (SIR) model \cite{first_sir_paper} where each individual in the population is associated with one epidemiological state at a time \cite{std_2_2_general_1,std_2_2_general_2}. Commonly, since different STDs have different recovery and re-infection patterns \cite{std_global_review}, models also adopted the SI, SIS, and SIRS types of models \cite{std_2_2_intro_1,sirs_model_std}. In order to further improve these models' accuracy, many properties such as gender and age are introduced to make the population more heterogeneous \cite{pop_split_models_1,pop_split_models_2,exp_2_2_2}. For instance, \cite{exp_2_2_1} proposed a SIR-based disease spread model with multiple transmission mechanisms, such as direct contact or vectors, and showed that the model captures the pandemic spread for large population sizes.

In addition, unlike airborne pathogens that infect individuals by close spatial proximity \cite{graph_2,airborne_sample_1,airborne_sample_2}, STDs are transmitted via sexual intercourse. Since sex is simply not random \cite{sex_not_random}, most models adopt a graph-based spatial component for the spread dynamics \cite{spatial_graph_3,spatial_graph_2,spatial_graph_1}. Regularly. the nodes of the graph representing the individuals in the population while the edges indicate one or more types of interaction between them \cite{graph_exp_1,graph_exp_2}. For example, \cite{exp_2_2_3} proposed a SIR-based model for STD spread on a bipartite random contact network. 

In this work, we follow this line of a STD pandemic spread model using an extended SIR model for the temporal component and a graph-based model for the spatial component.

\section{The Model}
\label{sec:model}

The proposed model consists of three interconnected components: a temporal component that describes a multi-pathogen STDs spread in the population; a spatial component that describes the interactions between individuals; and a dating app component that specifies how dating apps influence both the spatial and temporal dynamics. Each of these components, as well as the interactions between them, are detailed below. In addition, we propose an agent-based simulation implementation of this model to allow its \textit{in silico} investigation.

\subsection{Extend multi-pathogen SIS model}
\label{sec:temporal}
In order to capture the spread of a multi-pathogen STDs spread, we based our model on the work of \cite{msms}. However, this model is proposed for a generic multi-pathogen pandemic spread dynamics and does not capture four important processes in the context of multi-pathogen STDs spread. First, since many STDs have a significant exposure time \cite{std_e_1,std_e_2}, an Expose state (\(E\)) is introduced. Second, since individuals can recover from some STDs, and be re-infected later \cite{std_re_1,std_re_2}, we also introduce an immunity decay and re-infection dynamics to the model. Third, individuals can be infected simultaneously by multiple STDs \cite{std_multi}. Thus, we further extended the model to capture these dynamics. Finally, we remove the recovery (\(R\)) state as individuals do not develop long-term immunity to STDs, in general \cite{std_no_r_1,std_no_r_2}.

Formally, let us define a model that contains a finite population (\(P\)) of size \(n := |P|\) and their change over finite time \([t_0, t_f]\) such that \(t_f > t_0\). In addition, let us assume a set of disease-generating pathogens \(D\) such that \(|D| := k \in \mathbb{N}\). At each point in time, each individual is either susceptible (\(S\)), exposed (\(E\)), infected (\(I\)), or dead (\(D\)) from each of these pathogens. Hence, the epidemiological state of an individual is represented by a tuple \(\eta \in \{s, e, i, d\}^{k}\). For some reason, each individual belongs to a super-position epidemiological state where it is susceptible, exposed, and infected by a set of pathogens, \(s, e, i \subset D\), such that \(s \cap e \cap i = \emptyset \wedge s \cup e \cup i = D\) \cite{msms}. One can ignore the dead (\(d\)) state since if a single pathogen caused the death of the individual, the other states \(s, e,\) and \(i\) do not play any role in the individual's overall epidemiological state. 
 
As such, for each state, there are 12 processes that influence the number of individuals in each epidemiological state. First, individuals are born at some rate \(\alpha\). Second, individuals are infected by a pathogen \(j \in D\), becoming exposed to it with infection rate \(\beta\). Third, individuals that are exposed to a pathogen \(j\) become infectious at a rate \(\phi\). Fourth, individuals from the group \((s,e,i)\) are infected by a pathogen \(j \in s\) by animals from the same species, becoming exposed to it with an infection rate \(\beta\). Fifth, individuals from the group \((s,e,i)\) which are exposed to pathogen \(j \in e\) become infectious at a rate \(\phi\). Sixth, for each \(j \in i\) individuals from the group \((s,e,i)\) lose their immunity and become susceptible again to the pathogen \(j\) at a rate \(\psi\). Seventh, individuals from the group \((s,e,i)\) die due to their diseases at a rate \(\mu\). Finally, individuals are naturally dying at a rate \(\upsilon\), independent of the diseases they carry. These dynamics take the ordinary differential equations (ODEs) representation as follows:

\begin{equation}
    \begin{array}{l}
    \forall s, e, i: \frac{d P_{s, e, i}(t)}{d t} = \sum_{a, b, c | a \cap b \cap c = \emptyset \wedge a \cup b \cup c = D} \alpha_{a,b,c}P_{a,b,c} 

    + \sum_{j \in e} \beta_{s \cup j, e/j, i}^{s, e/j, i \cup j} P_{s \cup j, e/j, i} P_{s, e/j, i \cup j} \\ \\

    + \sum_{j \in i} \phi_{s, e \cup j, i/j} P_{s, e \cup j, i/j} 

    + \sum_{j \in s} \psi_{s/j, e, i \cup j} P_{s/j, e, i \cup j} 

    - \sum_{j \in s} \beta_{s, e, i}^{s, e/j, i \cup j} P_{s, e, i} P_{s, e/j, i \cup j, r}  \\ \\ 

    - \sum_{j \in e} \phi_{s, e, i} P_{s, e, i} 

    - \sum_{j \in i}  \mu_{s, e, i} P_{s, e, i} - \upsilon_{s, e, i} P_{s, e, i}
    
    \end{array} 
    \label{eq:pandemic_temporal}
\end{equation}

\begin{figure}[!ht]
    \centering
    \includegraphics[width=0.99\textwidth]{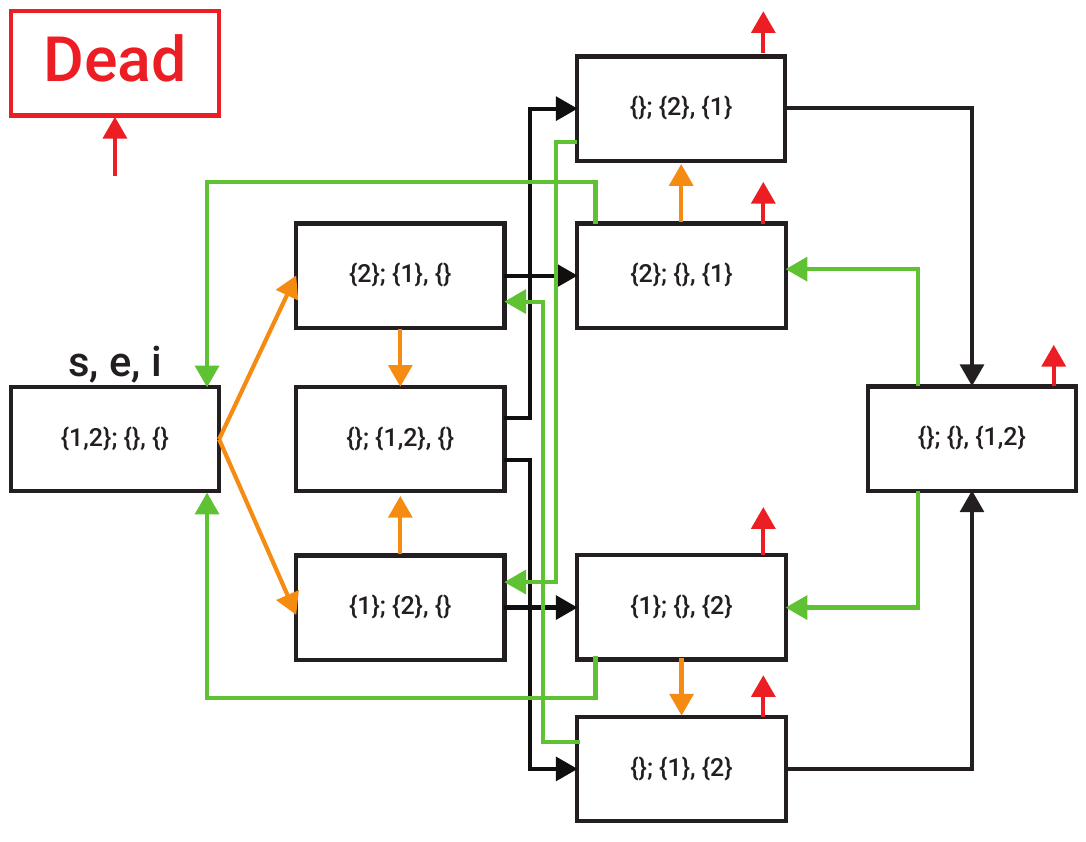}
    \caption{A schematic view of transition between disease stages, shown for \(k=2\). The red arrows indicate that from this stage, the individual might die from the disease. In a similar manner, the orange, black, and green arrows indicate exposure, infection, and recovery with immunity decay, respectively. }
    \label{fig:pandmeic_states_k_2}
\end{figure}

A schematic view of the epidemiological states of the model for the case of two pathogens (i.e., \(k=2\)) is shown in Fig. \ref{fig:pandmeic_states_k_2} where each box indicates the epidemiological state of the individual represented by the pathogens belongs to each of the \(s, e, i\) sets.

\subsection{Graph-based spatial interactions}
\label{sec:spatial}
Following the models proposed by \cite{sr_std_model} and \cite{teddy_chaos_paper}, for the proposed model, we adopted a two-layer graph-based spatial component. Formally, we consider a population of individuals, \(P\), that have two main types of interactions between them which are represented by two different \say{layers} of the interaction graph. The first layer, \(L_1\), represents steady partnerships among the individuals that resulted from socially accepted long-term sexual partnerships. In addition to these interactions, we assume a second type of interaction that corresponds to potential casual partnerships. These interactions become active with a probability, \(\xi \in [0, 1]\) when the individuals at both ends of the interactions are simultaneously seeking casual partners, aware of each other, and attracted to each other. This second \say{layer} of links is denoted by \(L_2\). By definition, \(L_1 \cap L_2 = \emptyset\). We assume that for each individual \(x \in \mathbb{P}\) in the population, there is a unique distribution function \(\delta_x(y)\) that obtains another individual in the population \(y \in \mathbb{P}\) and returns the probability that the individuals \(x\) and \(y\) would have a \(L_1\)-type interaction. In realistic social networks, each individual has a relatively small group of individuals with whom s/he has long-term sexual partnerships. In order to capture these dynamics in the infection graph, we assume \(\delta_x(y)\) follows a Poisson distribution with mean values \(\rho \in \mathbb{R}^+\). In addition, we assume that a \(L_1\) and \(L_2\) type edges can become \(L_2\) and \(L_1\) type edges, respectively, with probabilities \(\omega_1^2 \in [0, 1]\) and \(\omega_2^1 \in [0, 1]\) at each step in time.

\begin{figure}[!ht]
    \centering
    \includegraphics[width=0.7\textwidth]{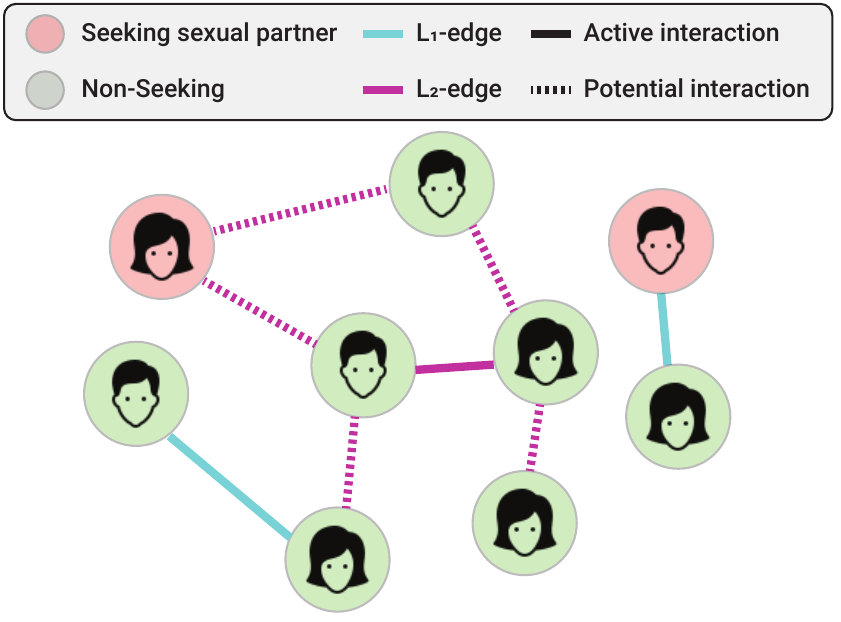}
    \caption{A schematic view of the interaction graph for a single point in time.}
    \label{fig:infection_graph}
\end{figure}

In addition, we assume that each individual is either seeking a sexual partner or not at any point in time, \(t\). When an individual seeks a partner, it first updates its \(L_2\) layer and choose randomly an individual from it, and establishes a casual partnership. Later, when one of the two individuals is no longer looking for a sexual partner state, the edge between the two nodes is removed. We assume node activation processes are independent Poisson processes \cite{sr_std_model}, where individual \(i\) seeks a sexual partner with rate \(\gamma^i_1 \in \mathbb{R}^+\), and if it is seeking for a sexual partner, it goes to the non-seeking state with rate \(\gamma^i_2 \in \mathbb{R}^+\). Due to the fact that the inverse of the transition rate is the expected value of transition time, if individual \(i\) seeks a sexual partner, it is expected to stay in this state for a period of time of length \((\gamma^i_2)^{-1} \in \mathbb{R}^+\). Moreover, individuals can interact in either protected or unprotected sexual interactions. If at least one of the sides prefers a protected interaction, it would be protected. Fig. \ref{fig:infection_graph} shows a schematic view of the interaction graph for a single point in time.

\subsection{Dating Apps dynamics}
\label{sec:dating_app}
Dating apps allow individuals in the population to meet more individuals than they would be able to by random encounters. More specifically, dating apps increase the rate at which both sides interact when they both seeking sexual partners as both individuals use dating applications only when they seeking sexual partners. Nonetheless, not all matches done in the dating application result in physical interaction \cite{dating_app_not_all_psycial}. The probability that a match on the dating app would result in a physical interaction depends on multiple factors. That said, one can simplify these into an abstract attractiveness level, \(b \in [0, 1]\) which each individual in the population has for any other individual in the population, which results in the population's attractiveness matrix, \(B \in [0, 1]^{n \times n}\). For simplicity, we assume that \(B\) is constant over time. 

Therefore, in order to further capture the heterogeneity of the population, for each individual we take into consideration its gender (\(g \in \{male, female\}\)) and age (\(a \in \mathbb{N}\)). These factors are used to determine the attractiveness level of an individual for other individuals according to their own gender and age as well as their preferences of gender and age in their sexual partners. We assume that gender and its preferences are constant over time while age and its preferences change identically over time. 

On top of that, dating apps shown empirically to be more popular in some social groups and their users' activity is also altering over time. To capture these dynamics, we assume that each individual in the population has a probability, \(d \in [0, 1]\), to use a dating app while seeking for sexual partner. Individuals who were successful in finding sexual partners using the dating app are more likely to re-use it by increasing their probability \(d\) by a factor of \(\delta_s \in [0, 1]\). On the other hand, individuals who were not successful in finding sexual partners in the dating app are more likely to use it less by decreasing their probability \(d\) by a factor of \(\delta_n \in [0, 1]\). 

\subsection{Assembling the components into a single framework using the agent-based simulation approach}
A powerful approach to implement this multi-component model into one framework is the agent-based simulation \cite{abs_1,abs_2,abs_3,abs_4}. Inspired by previous works \cite{abs_epi_1,teddy_ariel,abs_epi_2,alexi_security_games}, we formally define the model as a whole, denoted by \(M\), as follows. Let \(M\) be a tuple \((P, G)\) where \(P\) is a population of agents and \(G\) is the interaction graph between them. Let \(G := (P, E \subset P \times P \times \{1, 2\})\) be a two-layer connected graph where each node represents an individual in the population and the edge is a sexual interaction between two individuals. The individuals in the population are interacting in rounds, denoted by \(t \in [1, \dots, T]\) such that \(T<\infty\). Each individual in the population, \(p \in \mathbb{P}\), is represented by a timed finite state machine \cite{fsm}. An individual is described by a tuple \(p := (\eta, a, g, \mu, \theta, \gamma_1, \gamma_2, d, \delta_s, \delta_n, \omega_1^2, \omega_2^1, \zeta)\) where \(\eta\) is the agent's current epidemiological state, \(a\) is the agent's age, \(g\) is the agent's gender, \(\mu \in [0, 1]^n\) is the attractiveness level of all other individuals in the population according to the individual, \(\theta \in \{T, F\}\) indicates if the individual is currently seeking a sexual partner or not, \(\gamma_1\) and \(\gamma_2\) is the duration in which \(\theta\) changes between \(T \rightarrow F\) and \(F \rightarrow T\), respectively, \(d\) is the probability the individual would use a dating app while seeking for a sexual partner, \(\delta_s\) and \(\delta_n\) are the increase or decrease in \(d\) due either success or not in finding a sexual partner using the dating app, \(\omega_1^2, \omega_2^1\) are the probability that \(L_1\) and \(L_2\) type interactions would become \(L_2\) and \(L_1\) type interactions, respectively, and \(\zeta\) is a binary variable indicating if the agent wishes to participent in protected or unprotected sexual interaction. 

During the first round (\(t=1\)), the population (\(\mathbb{P}\)) is generated such that the individual's properties follow a pre-defined distribution. Moreover, the \(L_2\) layer in \(G\) is also generated. Afterward, at each round \(t\), each individual in the population, if seeking a sexual partner, can either try and increase the number of \(L_1\) type edges it has by using a dating app or not. Afterward, each individual chooses, at random, one of the \(L_1\) or \(L_2\) edges it has and interacts with the other individual. Similar to previous modeling attempts, we assume that all individuals interact in a single round. These interactions initiate some epidemiological dynamics, following Eq. (\ref{eq:pandemic_temporal}). Recall, that individuals with a susceptible status (\(S\)) have no immunity and are susceptible to infection by a pathogen \(i\). When an individual with an \(S\) status is exposed to the pathogen \(i\) through an interaction with an infected individual (\(I\) status), the individual is assigned with an exposed status \((E)\) with a probability \(\beta\) which corresponds to the \(\eta\) states of both individuals. Individuals with an \(E\)  status have the pathogen but are not yet contagious. The individual remains with an \(E\) status for \(\phi\) rounds, after which the individual is assigned with an infected status \((I)\), which makes her contagious to other individuals. After \(\gamma\) rounds, an infected individual transitions back to a susceptible status (\(S\)) or dead status \((D)\) with probabilities \((1-\psi)\) and \((\psi\)), respectively. Dead individuals are removed from the population (and the graph). In addition, at each step in time, new individuals are added to the population as they reach adulthood with a rate corresponding to the population size \(\alpha\).

\section{Experiment}
\label{sec:experiment}
In this section, we perform \textit{in silico} experiments based on the proposed model. Initially, we find from the literature realistic values for the model's parameters to obtain realistic realizations of the proposed model. Using this setup, we explore the influence of dating apps on the spread of STDs from three perspectives.

\subsection{Setup}
\label{sec:setup}
High-resolution and extensive epidemiological data are required to obtain a real-world realization of the proposed model. Unfortunately, currently, such data is unavailable in the public domain (to our best knowledge). Nonetheless, partial data about STD spread epidemics and general statistics about dating app usage are available in the literature \cite{std_decay_rate,birth_rate,death_rate}. Specifically, we focused on the three common STDs in the United States - Chlamydia, Gonorrhea, and Syphilis \cite{mortality_rate_stds}. In total, according to the Centers for Disease Control and Prevention, around 2.5 million cases of these diseases were reported during 2021 in the United States alone \footnote{We refer the interested reader to \url{https://www.cdc.gov/std/statistics/2021/default.htm} (visited 25th of September, 2023)}. On a more global scale, the World Health Organization (WHO) estimates 129, 82, and 7.1 million cases of Chlamydia, Gonorrhea, and Syphilis during 2020, respectively \footnote{The full report is available online \url{https://www.who.int/news-room/fact-sheets/detail/sexually-transmitted-infections-(stis)?gclid=CjwKCAjwpJWoBhA8EiwAHZFzfpYLqRvh-Tf2UNlypsUZz7s9frmif0aKHfur7LIw3kbzVsIQOa_oFhoCAYEQAvD_BwE} (visited 25th of September, 2023)}. In addition, to make the socio-demographic distribution realistic, we adopted the age and gender co-distribution from \cite{birth_rate}. In particular, for the average number of preeminent interactions, we computed the portion of officially married adults from the entire adult population, assuming only monogamic relationships. Table \ref{table:parameters} summarizes the proposed model's hyper-parameter values based on the available data from the literature, as stated in the \textbf{source} column. In particular, we choose to simulate a step in a time of one hour to balance the computational burden and the model's accuracy. Moreover, the population size range is chosen based on the estimation of sexually active adults in a small-medium US city.

\begin{table}[!ht]
\centering
\begin{tabular}{p{0.07\textwidth}p{0.45\textwidth}p{0.25\textwidth}p{0.1\textwidth}}
\hline \hline
\textbf{Symbol} & \textbf{Description} & \textbf{Default value} & \textbf{Source} \\ \hline \hline
   \(T\) & Number of simulation rounds (spanning over a year in the \(\Delta t\) used) [\(1\)] & 8760 & Assumed \\ 
   \(\Delta t\) & Simulation round's duration in time [\(t\)] & 1 hour & Assumed \\ 
   \(|P(0)|\) & The initial population size [\(1\)] & \([10^5, 10^6]\) & Assumed  \\
   \(k\)   & The number of pathogens [\(1\)] & \(3\) & Assumed \\
   \(\alpha\) & Birth rate in days [\(t^{-1}\)] & \(3.24 \cdot 10^{-5}\) & \cite{birth_rate} \\
   \(\upsilon\) & Natural death rate in days [\(t^{-1}\)]  & \(2.27 \cdot 10^{-5}\)  &  \cite{death_rate} \\
   \(\beta_{c}\) & Average Chlamydia infection rate [\(1\)]  & Protected - 2\%, Unprotected - 100\% & \cite{mortality_rate_stds} \\
   \(\beta_{g}\) & Average Gonorrhea infection rate [\(1\)] & Protected - 2\%, Unprotected - 100\% & \cite{mortality_rate_stds} \\
   \(\beta_{s}\) & Average Syphilis infection rate [\(1\)] &  Protected - 2\%, Unprotected - 100\% & \cite{mortality_rate_stds} \\
   \(\phi_{c}\) & Average Chlamydia exposure to infectious transformation rate  in days [\(t^{-1}\)] & 7-14 & \cite{NHS} \\
   \(\phi_{g}\) & Average Gonorrhea exposure to infectious transformation rate in days [\(t^{-1}\)] & 2-14 & \cite{cdc} \\
   \(\phi_{s}\) & Average Syphilis exposure to infectious transformation rate in days [\(t^{-1}\)] & 1-9 & \cite{cdc} \\
   \(\psi_{c}\) & Immunity decay rate for Chlamydia  in days [\(t^{-1}\)] & 0-1 & \cite{NHS} \\
   \(\psi_{g}\) & Immunity decay rate for Gonorrhea in days [\(t^{-1}\)] & 0-2 & \cite{std_decay_rate} \\
   \(\psi_{s}\) & Immunity decay rate for Syphilis in days [\(t^{-1}\)] & 0-2 & \cite{std_decay_rate} \\
   \(\gamma_{c}\) & Mortality rate due to Chlamydia [\(1\)] & \(1.8 \cdot 10^{-6}\) & \cite{mortality_rate_stds} \\
   \(\gamma_{g}\) & Mortality rate due to Gonorrhea [\(1\)] & 0 & \cite{mortality_rate_stds} \\
   \(\gamma_{s}\) & Mortality rate due to Syphilis [\(1\)] & 0 & \cite{mortality_rate_stds} \\ 
   \(\gamma_{1}\) & Sexual partner looking in hours [\(1\)] & \(N(0.72, 0.44)\) & \cite{app_base_data} \\ 
   \(\gamma_{2}\) & Sexual partner non-looking in hours & \(N(15.24, 6.73)\)  & \cite{app_base_data} \\ 
   \(d\) & Dating apps initial adoption rate [\(1\)]  & 0.38 & \cite{app_init_usage} \\ 
   \(\delta_s\) & Increase in personal usage probability of dating apps due to successful interaction using the app [\(1\)]  & 0.05 & Assumed \\ 
   \(\delta_n\) & Decrease in personal usage probability of dating apps due to successful interaction using the app  [\(1\)]  & 0.02 & Assumed \\ 
   \(\omega_1^2\) & A probability that casual interaction would become a preeminent interaction [\(1\)]  & 0.019 & \cite{apps_deep_con} \\ 
   \(\omega_2^1\) & A probability that preeminent interaction would become a casual interaction [\(1\)] & 0 & Assumed \\ 
   \(\mu\) & The average attractiveness distribution in the population [\(1\)]  & \(P(0.71)\)  & \cite{app_base_data} \\
   \(|P|/|L_1|\) & Average number of preeminent interactions [\(1\)] & 0.32 & Assumed \\
   & Portion of the population preferring protected sex [\(1\)] & 0.8 & Assumed \\ \hline
\end{tabular}
\caption{A summary of the proposed model's parameters and hyperparameters with their realistic value ranges. \(N(\mu, \sigma)\) indicates a normal distribution with a mean value of \(\mu\) and standard deviation of \(\sigma\). \(P(\lambda)\) indicates a Poisson distribution with a parameter \(\lambda\).}
\label{table:parameters}
\end{table}

Moreover, to assess the epidemic spread, we adopted the widely-used epidemiological metric - the average reproduction number (\(E[R_t]\)) which measures the number of secondarily infected individuals given the epidemic state at a given time \(t\) \cite{metric_paper_1,metric_paper_4,teddy_first,metric_paper_2,metric_paper_3}. Formally, \(R_t\) can be numerically approximated as follows:  \(R_t := \big ( I(t) - I(t-1) + S(t) - S(t-1)  \big ) / I(t-1)\), where \(I(t)\) and \(S(t)\) are the number of infected (by any pathogen) and recovered (and therefore susceptible again) individuals at time \(t\), respectively. Intuitively, the average reproduction number (\(E[R_t]\)) computes how many, on average, a single infected individual infects other individuals in a pre-defined and fixed duration (i.e., a step in time). 

\subsection{Results}
\label{sec:results}
Based on this setup, we conducted two main experiments as well as a sensitivity analysis for the model. First, we explore the influence of dating app adoption in the population on the STD spread dynamics. Second, we compare two scenarios of dating app usage - genuinely helping users to find stable relationships and promoting casual sexual encounters and further usage of the application. Finally, we explore the ability of dating apps to tackle the problem they (might) cause by introducing STD-aware and prevention policies\footnote{This question is inspired by recent such features by some dating apps: \url{https://www.statnews.com/2022/07/18/dating-apps-help-stop-spread-sexually-transmissible-infections/}}. 

\begin{figure}[!ht]
    \centering
    \includegraphics[width=0.99\textwidth]{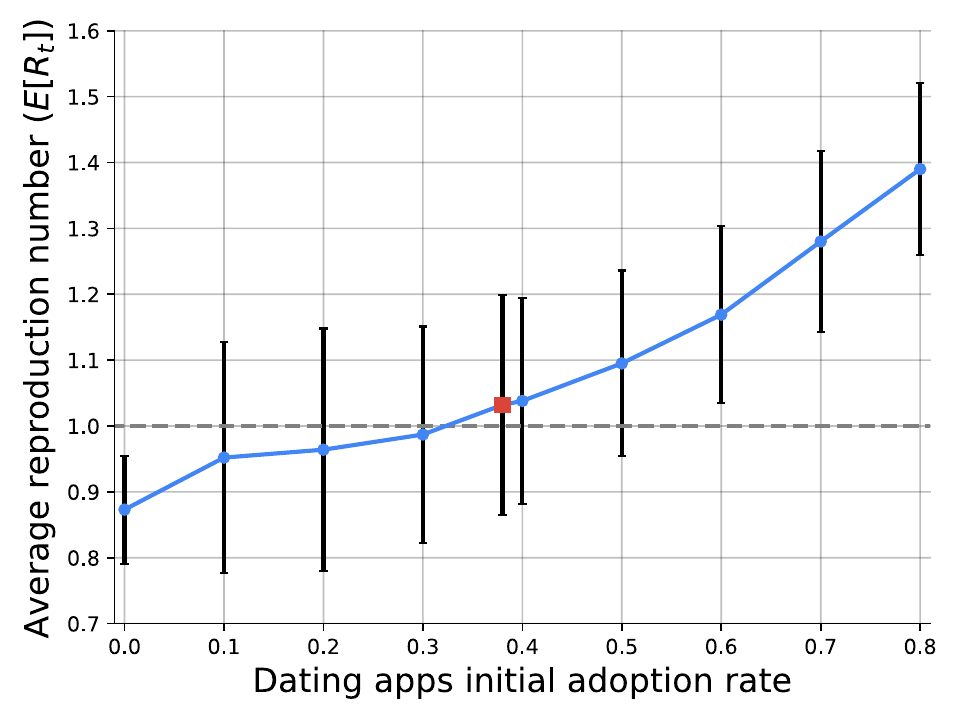}
    \caption{A comparison of the STD spread dynamics with different levels of dating app adoption. The results are shown as the mean \(\pm\) standard deviation of \(n=100\) simulation realizations. The case inferred from the historical data is marked by a red square while the other cases are marked by blue circles. The gray (dashed) line indicates \(E[R_t] = 1\) which is the epidemic outbreak threshold.}
    \label{fig:main}
\end{figure}

Fig. \ref{fig:main} presents the average reproduction number \((E[R_0])\) as a function of the dating app's initial adoption rate. The results are shown as the mean value of \(n=100\) simulation realizations and the error bars indicate the standard deviation of the sample. The case inferred from the historical data is marked by a red square while the other cases are marked by blue circles. The gray (dashed) line indicates \(E[R_t] = 1\) which is the epidemic outbreak threshold. One can notice that the increase in the dating apps' initial adoption rate caused a monotonic increase in the average reproduction number and therefore in the STD pandemic spread. Moreover, an increase of \(0.079\) in the average reproduction number occurs between no adoption and 0.1 adoption rate. Moreover, the average reproduction number increased rate is growing with the adoption rate, indicating a non-linear relationship between the two parameters. On the other hand, the standard deviations are (almost) monotonically decreased with respect to the adoption rate, excluding the case of no adoption. 

\begin{figure}[!ht]
    \centering
    \includegraphics[width=0.99\textwidth]{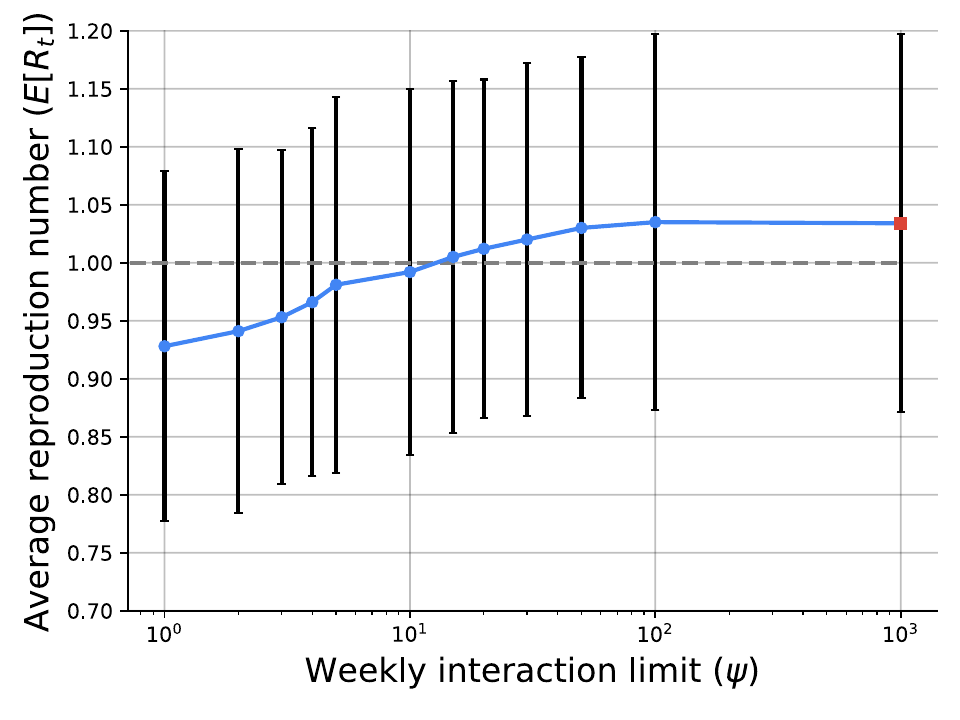}
    \caption{A comparison of the STD spread dynamics for two cases - genuinely helping users to find stable relationships and promoting casual sexual encounters and further usage of the application. The results are shown as the mean \(\pm\) standard deviation of \(n=100\) simulation realizations. The x-axis is presented in a logarithmic scale. The gray (dashed) line indicates \(E[R_t] = 1\) which is the epidemic outbreak threshold.}
    \label{fig:dating_app_style}
\end{figure}

After showing that dating apps adoption in its present form which encourages casual sexual interactions, we moved forward to investigate how changes in the application objective can influence STD spread. Namely, let us consider a scenario where dating apps limit one's ability to interact with other users over some period of time in order to motivate users to establish long-term relationships. Thus, we introduce a parameter, \(\psi \in \mathbf{N}\), which indicates how many interactions a user of the dating app is allowed to have in a week. For comparison, the present scenario assumes \(\psi \rightarrow \infty\) as no actual limit is present. Fig. \ref{fig:dating_app_style} shows the average reproduction number (\(E[R_t]\)) with respect to \(\psi\), demonstrating the STD spread dependency of how dating apps promoting casual sexual encounters and further usage of the application. The results are shown as the mean \(\pm\) standard deviation of \(n=100\) simulation realizations. One can notice a logarithmic relationship between the two parameters. Furthermore, with less than 10 possible interactions in the dating app, the STD epidemic is dying out as \(E[R_t] < 1\).  

\begin{figure}[!ht]
    \centering
    \includegraphics[width=0.99\textwidth]{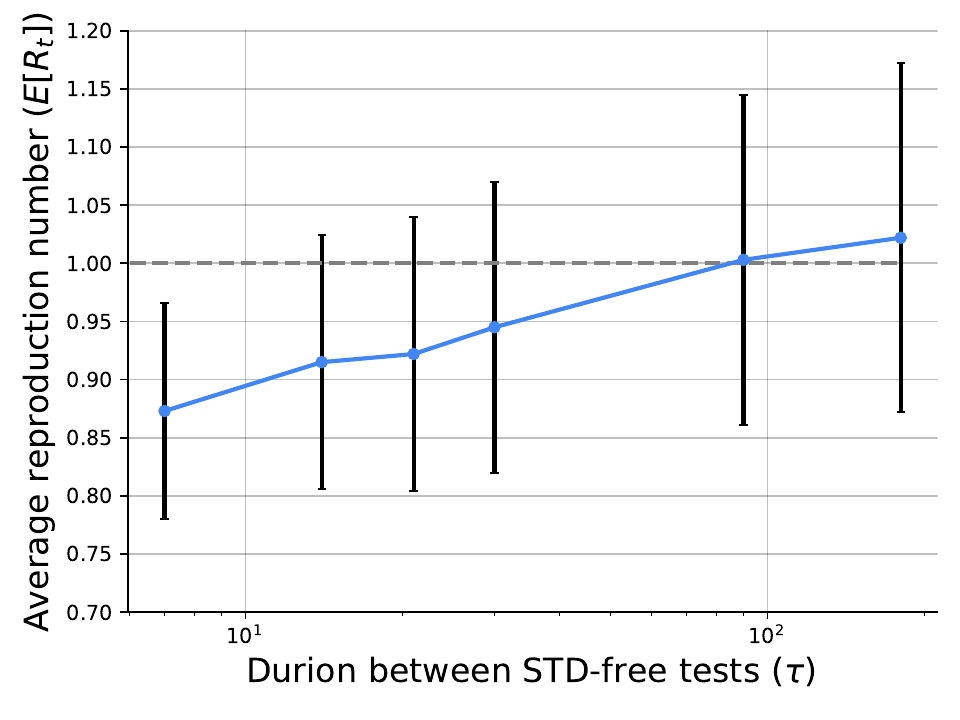}
    \caption{The average reproduction number (\(E[R_t]\)) with respect to the duration between two times a user has to prove it is STD-free \(\tau\). The results are shown as the mean \(\pm\) standard deviation of \(n=100\) simulation realizations. The x-axis is presented in a logarithmic scale. The gray (dashed) line indicates \(E[R_t] = 1\) which is the epidemic outbreak threshold.}
    \label{fig:dating_app_prevent}
\end{figure}

This outcome reveals that a restricted limit on users' usage of the dating app is applied, the lower (on average) the STD spread in the population. However, applying such a strategy is undesirable for dating apps that profit from users using the app. Hence, a more economically realistic option is the introduction of some enforcement mechanism that makes sure the dating app's users are not spreading STDs. One possible implementation of such an enforcement mechanism is to request users to present, periodically, an official document they are free of STDs. As such, users who are infected while required to present such a document would have to wait until they recover. To evaluate the performance of such an enforcement mechanism, we define, \(\tau \in \mathbb{N}\), the duration, in days, between two times a user needs to provide an STD-free document to the application. Fig. \ref{fig:dating_app_prevent} shows the average reproduction number (\(E[R_t]\)) with respect to \(\tau\) such that the values are presented as the mean \(\pm\) standard deviation of \(n=100\) simulation realizations. One can see a monotonic increase in both the mean and standard deviation of \(E[R_t]\) with respect to \(\tau\). 

\section{Discussion and Conclusion}
\label{sec:discussion}
In this study, we investigate the influence of dating apps on STDs spread in a population by applying a multi-pathogen epidemiological model. The proposed model is based on an extended SIR-based epidemiological model with a spatial component of infection graph, following a sequence of models designed and validated for STD spread analysis \cite{std_2_2_intro_1,std_2_2_intro_2,std_2_2_intro_3}. We implemented the proposed model as an agent-based simulation approach while taking into consideration a heterogeneous population and its usage of dating apps. We used historical STD epidemics as well as statistical data about dating app usage to obtain realistic realizations of the proposed model, capturing as closely as possible realistic spread dynamics in this context as previous models are shown to accurately capture similar epidemiological cases with only partial data \cite{multi_populations_4,multi_species_sir_model,sir_example_2,sir_example_9}.

Taken jointly, our results, as shown in Figs. \ref{fig:main}, \ref{fig:dating_app_style}, and \ref{fig:dating_app_prevent} show a simplistic and consistent outcome - larger usage and adoption of dating apps causes an increase in STD spread. This conclusion, while sounds trivial, has not been empirically explored yet. Previous studies show that more sexual interactions cause more STD spread and that dating apps cause more sexual interactions, on average \cite{dating_health_problem_2,more_sex_more_infection,dating_intro_1,intro_chat_2}. That said, only recently, a self-reported, retrospective, and relatively small sample size study was able to statistically associate the two processes \cite{dating_problem_example}. Thus, our result is the first to show a large-scale, while \textit{in silico}, connection between dating apps and STD spread. Moreover, we show (see Fig. \ref{fig:main}) that in its current form, more adoption of dating apps in the population would result in a polynomial increase in the average reproduction number of STDs which can quickly be developed into a large-size pandemic. Nonetheless, as presented by Figs. \ref{fig:dating_app_style} and \ref{fig:dating_app_prevent}, one can enforce some limitations upon dating apps to control the additional STD spread they cause. That said, such limitations would probably negatively influence these apps profits and therefore would not be initiated by their owner companies. Hence, a balance between the two can be achieved where users repeatedly use the dating app while also testing to prevent STD spread. Our analysis shows that every three-month test should prevent any STD outbreak over time, for example.

This research is not without limitations. First, in the proposed model we ignore the role healthcare services play in treating STDs in a direct manner which can alter the proposed results depending on the quality and quantity of this service to the population \cite{limit_1}. Second, we do not include a socially-aware factor that causes individuals who are aware they have STDs to make sure they do not infect others, as also requested by law in some countries \cite{limit_2}. Third, as evidence regarding the connection between porn and reduced sexual desire is gathered \cite{limit_3}, and in the scope of the digital world effect on STD spread, future work can also include the influence of porn. Namely, connecting the usage of porn to the duration of the non-seeking state of individuals in the proposed model. 

This study highlights the importance of taking into consideration the interactions occurring in the digital world as these influence the physical one, in the context of STD spreads via dating apps. Our model and simulation can be utilized to design and \textit{in silico} test various policies to tackle and control STD spread among the population.  

\section*{Declarations}
\subsection*{Funding}
This research does not receive any funding.

\subsection*{Conflicts of interest/Competing interests}
None.

\subsection*{Data availability}
The data that have been used in this study are publicly available in the referenced sources.

\subsection*{Acknowledgement}
The author wishes to thank Ariel Alexi and Ariel Fenster for helping with this study's administrative work. 

%\subsection*{Author Contribution}
%Teddy Lazebnik: Conceptualization, Resources, Software, Data curation, Formal Analysis, Validation, Investigation, Methodology, Visualization, Supervision, Writing - Original Draft, Writing - Review \& Editing. \\ 
 
\bibliography{biblio}
\bibliographystyle{unsrt}

\end{document}